\documentclass[11pt]{article}

\usepackage[a4paper,margin=2.25cm]{geometry}

\usepackage[utf8]{inputenc}

\usepackage[english]{babel}

\usepackage{amsfonts}

\usepackage{graphicx}

\usepackage{subfigure}

\usepackage{natbib}

\setcitestyle{authoryear,round,semicolon}

\usepackage{hyperref}

\title{Nowcasting the Portuguese GDP with Monthly Data}

\author{João B. Assunção\footnote{ORCID: \url{https://orcid.org/0000-0002-5576-3473}.}\\Pedro Afonso Fernandes\footnote{ORCID: \url{https://orcid.org/0000-0001-5762-5157}. Correspondence: Universidade Católica Portuguesa, Católica Lisbon School of Business \& Economics, Palma de Cima, Building 5, 4th floor, Room 5430, 1649-023 Lisboa, Portugal. Email: paf@ucp.pt.}\\\\Universidade Católica Portuguesa\\Católica Lisbon School of Business \& Economics\\Católica Lisbon Research Unit in Business \& Economics (CUBE)\\Católica Lisbon Forecasting Lab (NECEP)\\Portugal}

\date{\selectlanguage{english} \today}

\begin{document}

\maketitle

\selectlanguage{english}

\begin{abstract}

In this article, we present a method to forecast the Portuguese gross domestic product (GDP) in each current quarter (\emph{nowcasting}). It combines bridge equations of the real GDP on readily available monthly data like the Economic Sentiment Indicator (ESI), industrial production index, cement sales or exports and imports, with forecasts for the jagged missing values computed with the well-known Hodrick and Prescott (HP) filter. As shown, this simple multivariate approach can perform as well as a Targeted Diffusion Index (TDI) model and slightly better than the univariate Theta method in terms of out-of-sample mean errors.


\end{abstract}

\pagebreak

\section{Introduction}
\label{sec:intro}

Macroeconomic forecasting models are typically based on quarterly or annual data. Nevertheless, a lot of rich data are available on monthly, weekly or even daily basis including price indexes, economic sentiment/business climate indicators, industrial production, cement sales, car sales, unemployment, exports or imports among other indicators. In particular, monthly data already available for the current quarter could be combined with lagged quarterly data to produce better forecasts, either for the current quarter (\emph{nowcasting}) or for subsequent quarters, typically one or two periods ahead.

The estimation of current quarterly models from high frequency data was introduced in the late 1960's by the German-American economist Otto Eckstein [1927-1984] at Data Resources, Inc., a company integrated, meanwhile, in IHS Markit Ltd., now a part of S\&P Global. Though, it was the Nobel Laureate Lawrence Robert Klein [1920-2013] that formulated the two main approaches to the problem \citep{Klein1989}.

On the one hand, the analyst can consider the main entries on the expenditure side of national accounts and then establishes empirical \emph{bridge} equations by aggregating high-frequency (monthly) indicators into quarters and correlating those with quarterly components of the gross domestic product (GDP). On the other hand, the forecaster can collect as much high frequency data as are available in the current quarter and then she extracts the leading principal components of the quarterly averages of these several indicators and regresses the GDP on them.

Bridge equation methods, also known as ``tracking models'' \citep{Higgins2014}, explores the interrelationships between expenditure (or income) components of GDP and monthly available data. For example, the private consumption of durable goods might be highly correlated with car sales, and other relationships could be found for national quarterly accounts (NQA) entries like gross private residential and non-residential investment, inventory changes, government expenditure and exports and imports of goods and services \citep{Klein1989}.

Where possible, bridge equations are estimated from monthly data on both high frequency indicators and NQA components. However, national accounts are not reported monthly in most cases, so quarterly bridge equations must be build by aggregating or averaging the monthly indicators into quarters. Typically, end-of-sample jagged values for those indicators are forecasted with monthly univariate models, either to complete the current quarter or to forecast one or two quarters beyond.

The first approach, originally proposed by \citet{Klein1989}, builds separate bridge equations for nominal GDP components and price deflators, noting that real expenditure estimates can be obtained by dividing the estimated nominal value of each NQA component by an appropriate deflator, which may be highly correlated with consumer price index. Typically, a bridge equation relates each NQA entry with the quarterly averages of one or two highly correlated variables. Finally, a national accounting entity is used to sum up the estimated (real) NQA components into GDP.

The second approach to nowcasting is founded mainly on the contributions of \citet{Stock1989,Stock2002,Stock2011}. The premise of their models is that a few latent dynamic factors drive the comovements of a time series like GDP, which is also affected by zero-mean idiosyncratic disturbances. These disturbances arise from measurement error and from specific features of the data. The latent factors follow a time series process, typically a vector autoregression (VAR). This kind of ``medium data'' or ``data-rich'' forecasting methods without expenditure components \citep{Higgins2014} can cover more than 200 monthly macroeconomic indicators like the application of \citet{Giannone2008}.

In Portugal, the central bank (Banco de Portugal, BdP) has a long experience in forecasting GDP using dynamic factor models with autoregressive components, the so-called Diffusion Index (DI) approach \citep{Stock2002}. In this kind of models, a large number of predictors is summarized using a small number of indexes constructed by principal component analysis, and a dynamic factor model serves as the statistical framework for the estimation of the indexes and forecasting. As stressed by \citet{Dias2015}, these models requires the previous determination of the factors which reflect  the top-ranked principal components, that is, the ones that encompass the largest share of the common comovement in the dataset; all other lower-ranked factors are entirely disregarded independently of their possible informational content for forecasting the variable of interest, typically, the GDP growth.

In order to avoid this limitation of DI approach, \citet{Dias2010} develop the targeting principle discussed by \citet{Bai2008}. Their procedure considers a synthetic regressor which is computed as a linear combination of all the factors of the dataset weighted such that each factor takes into account both the relative size of the variance captured by it and its correlation with the variable of interest at the relevant forecast horizon. Recent findings for the Portuguese real GDP growth \citep{Dias2016} suggest that this Targeted Diffusion Index (TDI) approach could reduce the out-of-sample mean squared error (MSE) by 63\% using a simple autoregressive model as benchmark, while the gain with a DI model is about 50\%.

The combination of the tracking and data-rich approaches is also possible. In fact, bridge equations can include common factors and may be applied either to the aggregate GDP or to its components. The GDPNow model from the Federal Reserve Bank of Atlanta \citep{Higgins2014} is a good attempt in that direction, despite its complexity and detail level.

In this paper, we present an approach that combines bridge equations of real GDP on several covariates available on a monthly basis, including coincident indicators, with forecasts for the jagged missing values computed with the well-known \citet{Hodrick1997} filter. As described below, this multivariate approach can perform as well as the TDI model developed by the Portuguese central bank, and slightly better than the univariate Theta method.

\section{Methodology}
\label{sec:method}

Our general methodology is similar to the approach proposed by \citet{Miller1996}. Firstly, we use a simple model to predict each relevant seasonally adjusted monthly series $Y_{t:i}$  for the current quarter $t$, where $i = 1, 2, 3$ is the number of months of data available from it. For example, when two months of data are available, the forecast for the third month of $t$ is

\begin{equation}
	\hat{Y}_{t:3} \equiv Y_{t:2} + g_{t:2}
	\label{eq:fcast}
\end{equation}

\noindent where $g_{t:2} \equiv X_{t:2} - X_{t:1}$ is the first difference of the trend $X_{s:i}$ of the series $Y_{s:i}$, $s=1, .., t$, $i = 1, 2, 3$, which is chosen to minimize either the sum of the square residuals $\varepsilon_{s:i}=Y_{s:i}-X_{s:i}$ or its smoothness \citep{Hodrick1997}:

\begin{equation}
	\textrm{min} \left\{  \sum_{s,i}^{} \varepsilon_{s:i}^2 + \lambda \sum_{s,i}^{} \left[ (X_{s:i} - X_{s:i-1}) - (X_{s:i-1}- X_{s:i-2})\right]^2  \right\}
	\label{eq:HPobjective}
\end{equation}

\noindent where $\lambda = 14400 $ is a penalty for the square of the acceleration (second difference) of the trend.

For raw series  $Y_{t:i}$ particularly noisy such as the industrial production index (\emph{IPI}), cement sales (\emph{CEM}) or card transactions in point of sales or automated teller machines (\emph{ATM}), we compute the following alternative moving average (MA) forecast for the same example:

\begin{equation}
	\tilde{Y}_{t:3} \equiv \frac{Y_{t:2}+Y_{t:1}+Y_{t-1:3}}{3} + \frac{g_{t:2}+g_{t:1}+g_{t-1:3}}{3} \times 2
	\label{eq:MAfcast}
\end{equation}

The number of months $i$ of data available depends upon the time series and the day of the current quarter. For instance, at day 60, two months of the European Commission's Economic Sentiment Indicator (\emph{ESI}) are available, but only the first month of the industrial production index is ready to use, that is, the IPI is delayed one month. The table \ref{tab:nowdata} indicates the selected series available at days 0, 30, 60 and 90 of the current quarter $t$, as well as at the day 10 of the next quarter $t+1$, here designated by day 100 for convenience.

\begin{table}[h]
	\centering
	\caption{Number of months of data available at the days 0, 30, 60, 90 and 100 of the current quarter $t$}
	\label{tab:nowdata}
	\begin{tabular}{lcccccc}
		\\[-2ex]\hline
		\multicolumn{2}{}{} & \multicolumn{5}{c}{Day of the current quarter $t$} \\
		\cline{3-7}
		Selected time series				& Source				& 0		& 30	& 60	& 90	& 100 \\
		\hline
		Economic Sentiment Indicator (ESI)	& European Comission	& t-1:3	& t:1	& t:2	& t:3	& t:3 \\
		Economic Climate Indicator (ICE)	& Statistics Portugal	& t-1:3	& t:1	& t:2	& t:3	& t:3 \\
		Industrial Prod. Index (IPI)		& Statistics Portugal	& t-1:2	& t-1:3	& t:1	& t:2	& t:2 \\
		Cement sales (CEM)					& Ministry of Finance	& t-1:2	& t-1:3	& t:1	& t:2	& t:3 \\
		Car sales	 (CAR)					& ACAP					& t-1:2	& t-1:3	& t:1	& t:2	& t:3 \\
		Card transactions (ATM)				& Banco de Portugal		& t-1:2	& t-1:3	& t:1	& t:2	& t:2 \\
		Exports of goods \& services (EXGS)	& Banco de Portugal		& t-1:1	& t-1:2	& t-1:3	& t:1	& t:1 \\
		Imports goods \& services (IMGS)	& Banco de Portugal		& t-1:1	& t-1:2	& t-1:3	& t:1	& t:1 \\
		Exports of goods (EXG)				& Statistics Portugal	& t-1:1	& t-1:2	& t-1:3	& t:1	& t:2 \\
		Imports of goods (IMG)				& Statistics Portugal	& t-1:1	& t-1:2	& t-1:3	& t:1	& t:2 \\
		Consumer Price Index (CPI)			& Statistics Portugal	& t-1:2	& t-1:3	& t:1	& t:2	& t:3 \\	
		Brent Oil Price (OIL)				& FRED \& Bloomberg		& t-1:3	& t:1	& t:2	& t:3	& t:3 \\
		Euro-coin (CEPR)					& Banca d'Italia \& CEPR& t-1:3	& t:1	& t:2	& t:3	& t:3 \\		
		\hline
	\end{tabular}
\end{table}

The monthly series of exports (\emph{EXGS}) and imports (\emph{IMGS}) of goods and services are delayed two months with only one month available, even 10 days after the end of the current quarter. However, two months of the series of exports and imports of goods (without services, denoted respectively \emph{EXG} and \emph{IMG}) are available at that time. Thus, we perform a simple linear regression to forecast an additional figure $\widehat{EXGS}_{t:2}$ (or $\widehat{IMGS}_{t:2}$) for the exports (imports) of goods and services:

\begin{equation}
	\left\{
	\begin{array}{l}
		\widehat{EXGS}_{t:2} = EXGS_{t-4:2} \times (1 + \widehat{exgs}_{t:2}/100) \\
		\widehat{exgs}_{t:2} = \hat{\alpha} + \hat{\beta} exg_{t:2}
	\end{array}
	\right .
	\label{eq:XMfcast}
\end{equation}

\noindent where the lower cases $exgs$ and $exg$ are the percentage growth rates compared to the same quarter of previous year for the corresponding upper case variables, namely, $exg_{t:2} = (EXG_{t:2}/EXG_{t-4:2}-1) \times 100$, where $EXG_{t:2}$ is the data on exports of goods, available at the day 100 of the current quarter $t$, and $\hat{\alpha}$ and $\hat{\beta}$ are the ordinary least squares (OLS) coefficients on these year-over-year (y-o-y) changes in order to avoid serial correlation and residual seasonality issues. Similar equations were defined for imports. As usual, the third month of exports (or imports) of goods and services is estimated using formula (\ref{eq:MAfcast}) and the Hodrick-Prescott filter with the additional figure $\widehat{EXGS}_{t:2}$ (or $\widehat{IMGS}_{t:2}$).

Secondly, we average monthly data and forecasts for each quarter. This procedure is necessary because our variable of interest (\emph{GDP}) is issued quarterly. So, in the same example of two months of data available in quarter $t$, we compute:

\begin{equation}
	\hat{Y}_{t} = \frac{Y_{t:1}+Y_{t:2}+ \hat{Y}_{t:3}}{3}
	\label{eq:MA3}
\end{equation}

\noindent where the generic $\hat{Y}_{t:3}$ can be replaced by $\tilde{Y}_{t:3}$ for noisy series. The exports and imports of goods and services are still a special case in the sense that their quarterly figures are estimated using a specific multiple regression model which takes into account, as dependent variable, the exports \emph{EXP} (or imports \emph{IMP}) in volume provided by the NQA\footnote{Exports and imports are available only 60 days after the end of the respective quarter.} and, as independent variables, the monthly exports (or imports) of goods and services previously averaged with formula (\ref{eq:MA3}), the lagged price deflator (\emph{DEF}) of exports (or imports) and the Brent oil price (\emph{OIL}). This procedure is required because the monthly series of exports and imports of goods and services are provided in current prices, that is, in (nominal) values instead of (real) volumes. As in (\ref{eq:XMfcast}), the lower case points out the percentage change over the same quarter of the previous year (y-o-y) of the correspondingly level variable in upper case:

\begin{equation}
	\left\{
	\begin{array}{l}
		\widehat{EXP}_{t} = EXP_{t-4} \times (1 + \widehat{exp}_{t}/100) \\
		\widehat{exp}_{t} = \hat{\alpha}_0 + \hat{\alpha}_1 exgs_{t} + \hat{\alpha}_2 def_{t-1}  + \hat{\alpha}_3 oil_{t}\\
	\end{array}
	\right .
	\label{eq:XMfcast2}
\end{equation}

Thirdly, the current quarter's GDP growth ($gdp_t$) is predicted using several bridge equations which explore different combinations of the monthly time series listed in table \ref{tab:nowdata}. Here, we describe six linear models from a pool of more than thirty models currently in use.

The first of these models, denoted with the superscript $^{(1)}$, includes, as independent variables, the sum of the last three quarter-over-quarter (q-o-q) percentage growth rates of real GDP ($sum_{t-1}$)\footnote{Before 2020, the last GDP growth rate was issued only 45 days after the end of the respective quarter, so it cannot be summed up to nowcast the current (next) quarter until then. Thus, the forecasts at days zero and 30 of the current quarter $t$ were made using the sum of the last two available q-o-q changes of GDP, concerned with periods $t-2$ and $t-3$.}, the quarterly averages of the Economic Sentiment Indicator ($\widehat{ESI}_t$) previously regularized (by subtracting the historical average 100) and Economic Climate Indicator ($\widehat{ICE}_t$), as well the y-o-y changes of industrial production index ($\widehat{ipi}_t$) and cement sales ($\widehat{cem}_t$):

\begin{equation}
\widehat{gdp}_{t}^{(1)} = \hat{\alpha}_0 + \hat{\alpha}_1 sum_{t-1} + \hat{\alpha}_2 \widehat{ESI}_{t} + \hat{\alpha}_3 \widehat{ICE}_{t} +\hat{\alpha}_4 \widehat{ipi}_{t}  + \hat{\alpha}_5 \widehat{cem}_{t}\\
\label{eq:mod1}	
\end{equation}

\noindent where $\hat{\alpha}_0, \dots, \hat{\alpha}_5$ are the coefficients estimated with ordinary least squares (OLS). 

The second model adds the international trade, that is, the real y-o-y changes of exports and imports in addition to the independent variables already considered in equation (\ref{eq:mod1}):

\begin{equation}
\widehat{gdp}_{t}^{(2)} = \hat{\alpha}_0 + \hat{\alpha}_1 sum_{t-1} + \hat{\alpha}_2 \widehat{ESI}_{t} + \hat{\alpha}_3 \widehat{ICE}_{t} +\hat{\alpha}_4 \widehat{ipi}_{t}  + \hat{\alpha}_5 \widehat{cem}_{t} + \hat{\alpha}_6 \widehat{exp}_{t} + \hat{\alpha} \widehat{imp}_{t}\\
\label{eq:mod2}	
\end{equation}

The third and fourth models replaces \emph{ICE} in equation (\ref{eq:mod1})  with the y-o-y changes of car sales and card transactions, respectively. Cards transactions are expressed in real terms, that is, they were previously deflated with the consumer price index (\emph{CPI}).

\begin{equation}
\widehat{gdp}_{t}^{(3)} = \hat{\alpha}_0 + \hat{\alpha}_1 sum_{t-1} + \hat{\alpha}_2 \widehat{ESI}_{t} + \hat{\alpha}_3 \widehat{car}_{t} +\hat{\alpha}_4 \widehat{ipi}_{t}  + \hat{\alpha}_5 \widehat{cem}_{t}\\
\label{eq:mod3}	
\end{equation}

\begin{equation}
\widehat{gdp}_{t}^{(4)} = \hat{\alpha}_0 + \hat{\alpha}_1 sum_{t-1} + \hat{\alpha}_2 \widehat{ESI}_{t} + \hat{\alpha}_3 \widehat{atm}_{t} +\hat{\alpha}_4 \widehat{ipi}_{t}  + \hat{\alpha}_5 \widehat{cem}_{t}\\
\label{eq:mod4}	
\end{equation}

 The fifth model is based in the last model, but includes the international trade:

\begin{equation}
\widehat{gdp}_{t}^{(5)} = \hat{\alpha}_0 + \hat{\alpha}_1 sum_{t-1} + \hat{\alpha}_2 \widehat{ESI}_{t} + \hat{\alpha}_3 \widehat{atm}_{t} +\hat{\alpha}_4 \widehat{ipi}_{t}  + \hat{\alpha}_5 \widehat{cem}_{t} + \hat{\alpha}a_6 \widehat{exp}_{t} + \hat{\alpha}_7 \widehat{imp}_{t}\\
\label{eq:mod5}	
\end{equation}

Finally, the model 6 replaces \emph{exp} and \emph{imp} in equation (\ref{eq:mod2})  with the Euro-coin Real Time Indicator of the Euro Area Economy (\emph{CEPR}), an alternative measure of the external outlook:

\begin{equation}
\widehat{gdp}_{t}^{(6)} = \hat{\alpha}_0 + \hat{\alpha}_1 sum_{t-1} + \hat{\alpha}_2 \widehat{ESI}_{t} + \hat{\alpha}_3 \widehat{ICE}_{t} +\hat{\alpha}_4 \widehat{ipi}_{t}  + \hat{\alpha}_5 \widehat{cem}_{t} + \hat{\alpha}_6 \widehat{CEPR}_{t}\\
\label{eq:mod6}	
\end{equation}

For each model $j = 1, \dots, 6$, we compute an alternative GDP growth estimate with a simple error correction mechanism that incorporates the last observed error $\epsilon_{t-1}^{(j)}$:

\begin{equation}
\widehat{gdp}_{t}^{(jc)} = \widehat{gdp}_{t}^{(j)} + \epsilon_{t-1}^{(j)} = \widehat{gdp}_{t}^{(j)} + \left( gdp_{t-1} - \widehat{gdp}_{t-1}^{(j)} \right) \\
\label{eq:error_corr}	
\end{equation}

Then, we apply the median operator X to obtain a consensus among the corrected and uncorrected (simple) forecasts for each model $j$ and also for the six models such that

\begin{equation}
\widehat{gdp}_{t}^{(c)} = \textrm{X} \left( \widehat{gdp}_{t}^{(1)}, \widehat{gdp}_{t}^{(1c)}, \dots, \widehat{gdp}_{t}^{(6)}, \widehat{gdp}_{t}^{(6c)} \right)\\
\label{eq:consensus_wide}	
\end{equation}

We also perform a consensus among the simple forecasts without error correction:

\begin{equation}
\widehat{gdp}_{t} = \textrm{X} \left( \widehat{gdp}_{t}^{(1)}, \dots, \widehat{gdp}_{t}^{(6)} \right)\\
\label{eq:consensus}	
\end{equation}

As benchmark, we calculated one period and two periods look ahead forecasts with the Theta method proposed by \citet{Assimakopoulos2000} and explained by \citet{Hyndman2001}. Briefly, this method starts with the estimation of the following regression in level:

\begin{equation}
(1  - \theta) GDP_{t} = \hat{a}_{\theta} + \hat{b}_{\theta} (t-1) \\
\label{eq:theta}	
\end{equation}

\noindent for $\theta = 0$ and $\theta = 2$, where $t=1,...,n$ is the time index. Note that, when $\theta = 0$, $\hat{a}_0$ and $\hat{b}_0$ are simply the parameters of the linear time trend fitted to the GDP quarterly series. Then, for each value of $\theta$, a new series $GDP_{t,\theta}$ is constructed by doing:

\begin{equation}
GDP_{t,\theta} = \hat{a}_{\theta} + \hat{b}_{\theta} (t-1) + \theta \times GDP_{t}.\\
\label{eq:newtheta}	
\end{equation}

\noindent The $h$-step ahead forecast is obtained by averaging the GDP forecasts for $\theta=0,2$:

\begin{equation}
\widehat{GDP}_{t}(h) = \frac{\widehat{GDP}_{t,0}(h) + \widehat{GDP}_{t,2}(h)}{2}\\
\label{eq:thetafcast}	
\end{equation}

\noindent where $\widehat{GDP}_{t,0}(h)$ is obtained by extrapolating the linear time trend:

\begin{equation}
\widehat{GDP}_{t,0}(h) = \hat{a}_0 + \hat{b}_0 (t+h-1)\\
\label{eq:theta0fcast}	
\end{equation}

\noindent and $\widehat{GDP}_{t,2}(h)$ is obtained using simple exponential smoothing (SES) on series $\{GDP_{t,2}\}$:

\begin{equation}
\widehat{GDP}_{t,2}(h) = \gamma GDP_{t,2} + (1-\gamma) \widehat{GDP}_{t-1,2} \\
\label{eq:theta2fcast}	
\end{equation}

\noindent where $\widehat{GDP}_{1,2}=GDP_{1,2}$ (starting value) and $\gamma=0.3$ (smoothing parameter). Thus, the SES forecasts are equivalent for all $h$ \citep{Hyndman2001}.

\section{Main findings}
\label{sec:findings}

In this application, we considered the quarter-over-quarter (chain) percentage changes of real GDP, exports and imports provided by Statistics Portugal (INE) since the first quarter of 1996 till the fourth quarter of 2019 (total of 96 observations), fully available 60 days after the end of the respective quarter, and complemented by the monthly data indicated in table \ref{tab:nowdata} (above).\footnote{Most recent data were not considered in this paper because the COVID-19 pandemic and lockdowns had originated a structural break in the GDP series, namely, for Portugal in the first quarter of 2020 that had required a different forecasting strategy and method, based on differences in levels instead of growth rates.}

An out-of-sample rolling forecasting exercise was performed since the first quarter of 2002 to assess the relative performance of the described nowcasting models. Each current quarter growth was estimated with the data available at day 0, 30, 60, 90 and 100. For days 0 and 30, the benchmark is the 2-steps ahead forecast given by the Theta method, recalling that the GDP change of the previous quarter was available only at day 45 of the current quarter (and the exports and imports at day 60) before 2020. For days 60, 90 and 100, we used the one period look ahead Theta forecast as benchmark.

The estimated coefficients for the full sample (1996Q1 - 2019Q4) are presented in table \ref{tab:coef_2019}. In general, these coefficients are 1\% or 5\% statistically significant through the six models with few exceptions concerned, namely, with industrial production index (IPI). Particularly relevant is the coefficient associated with the last q-o-q changes of real GDP (sum), suggesting the auto-correlated nature of the dependent variable. The six models have high, significant F statistics and adjusted R$^{2}$ above 90\%, specially the models 2 and 5 with exports and imports.

Truncating the data may have a limited impact on these estimates. In fact, the coefficients presented in table \ref{tab:coef_2005} for roughly an half of the sample (40 observations from 1996Q1 to 2005Q4) are similar from those condensed in table \ref{tab:coef_2019}. Nevertheless, IPI loosened its significance in all models and, sometimes, the Economic Sentiment Indicator (ESI) and the CEPR indicator.

\begin{figure}[!h]
	\centering
	\subfigure[Model 1]{\includegraphics[width=8cm]{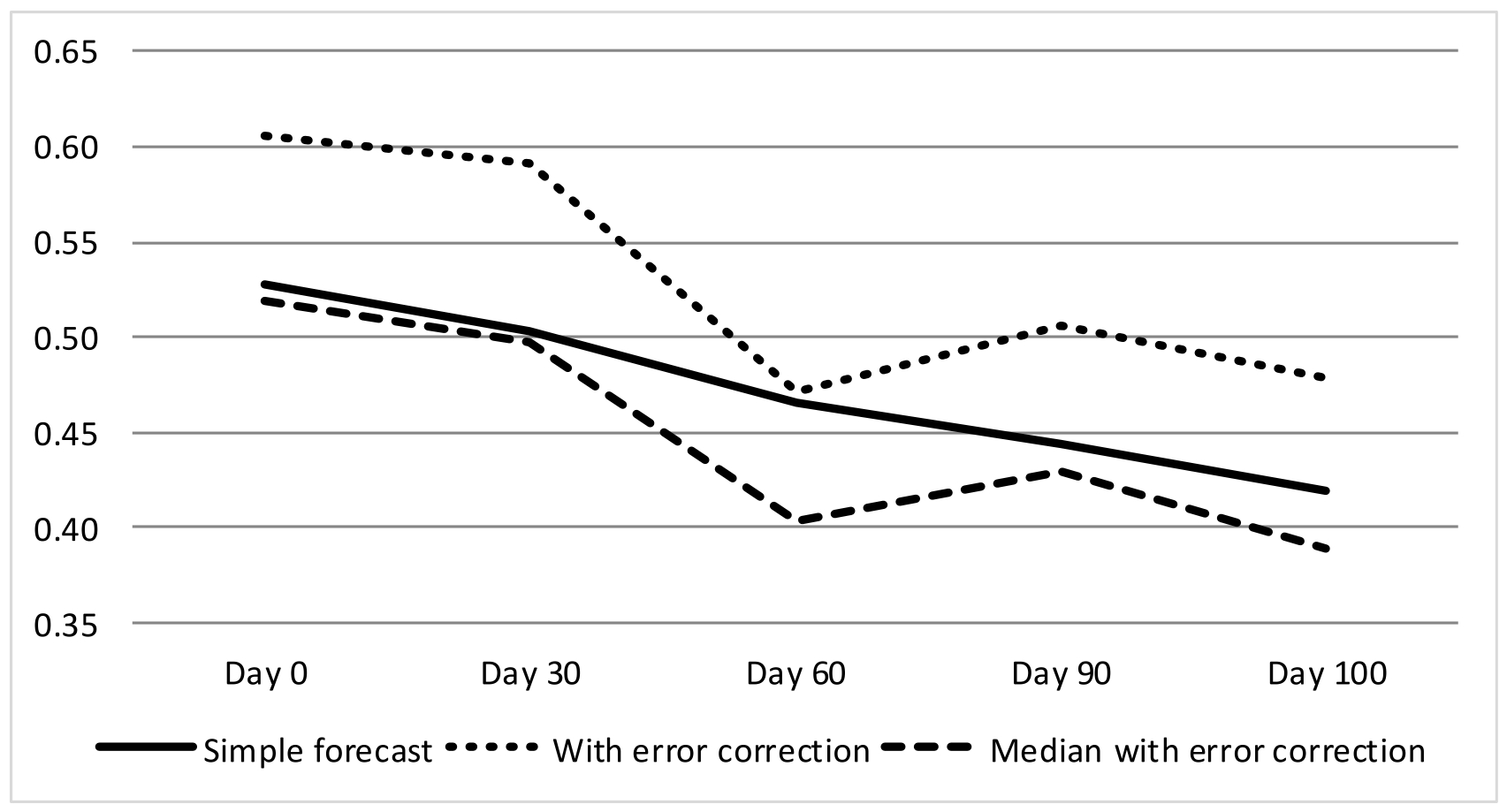}}
	\subfigure[Model 2]{\includegraphics[width=8cm]{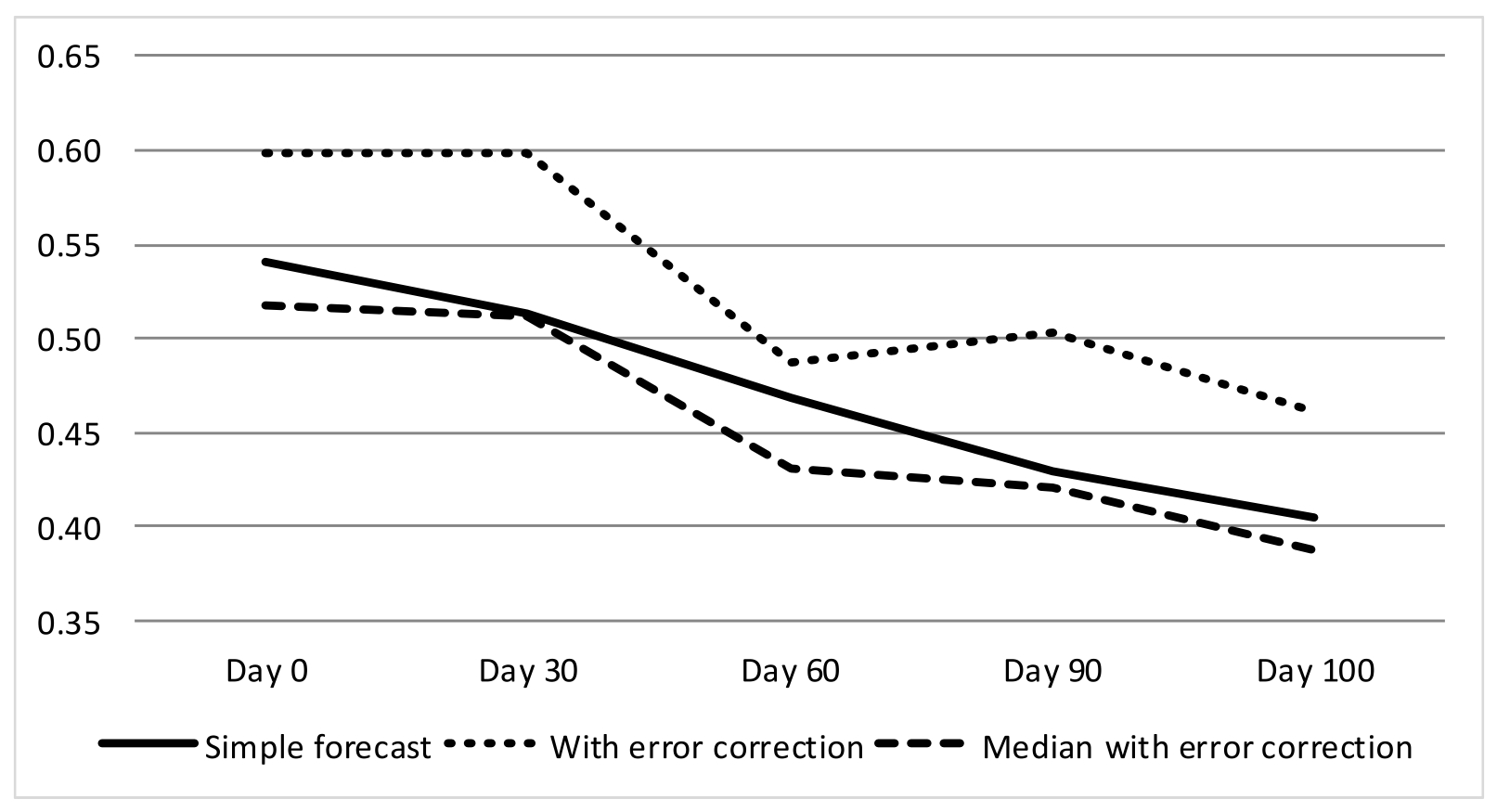}}
	\caption{Mean absolute out-of-sample forecast error of models 1 and 2, with and without last error correction, progressively updated with more data over the current quarter (percentage points, 2002Q1-2019Q4)}
	\label{fig:mod1_2}
\end{figure}

\begin{figure}[!h]
	\centering
	\subfigure[Model 3]{\includegraphics[width=8cm]{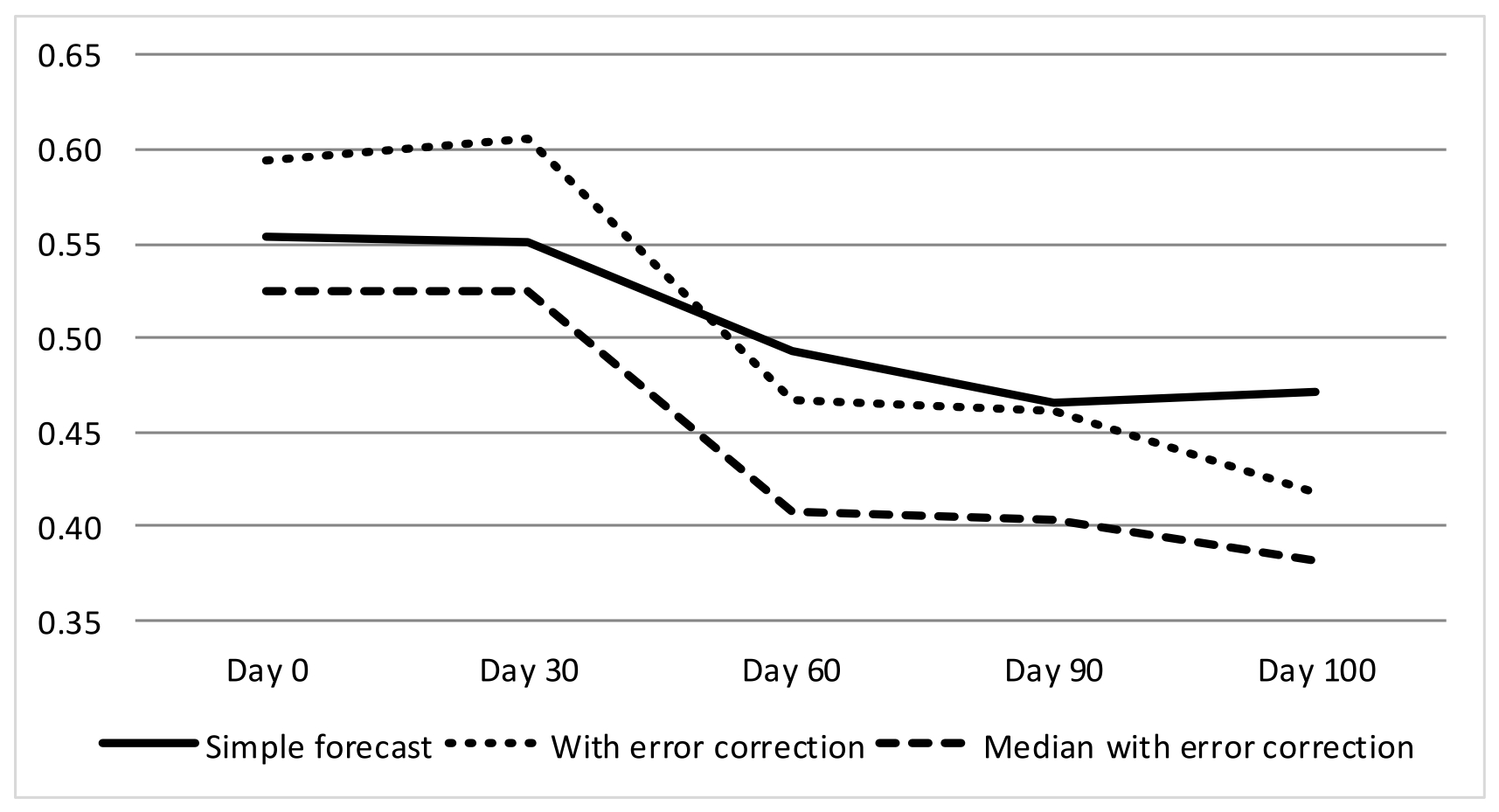}}
	\subfigure[Model 4]{\includegraphics[width=8cm]{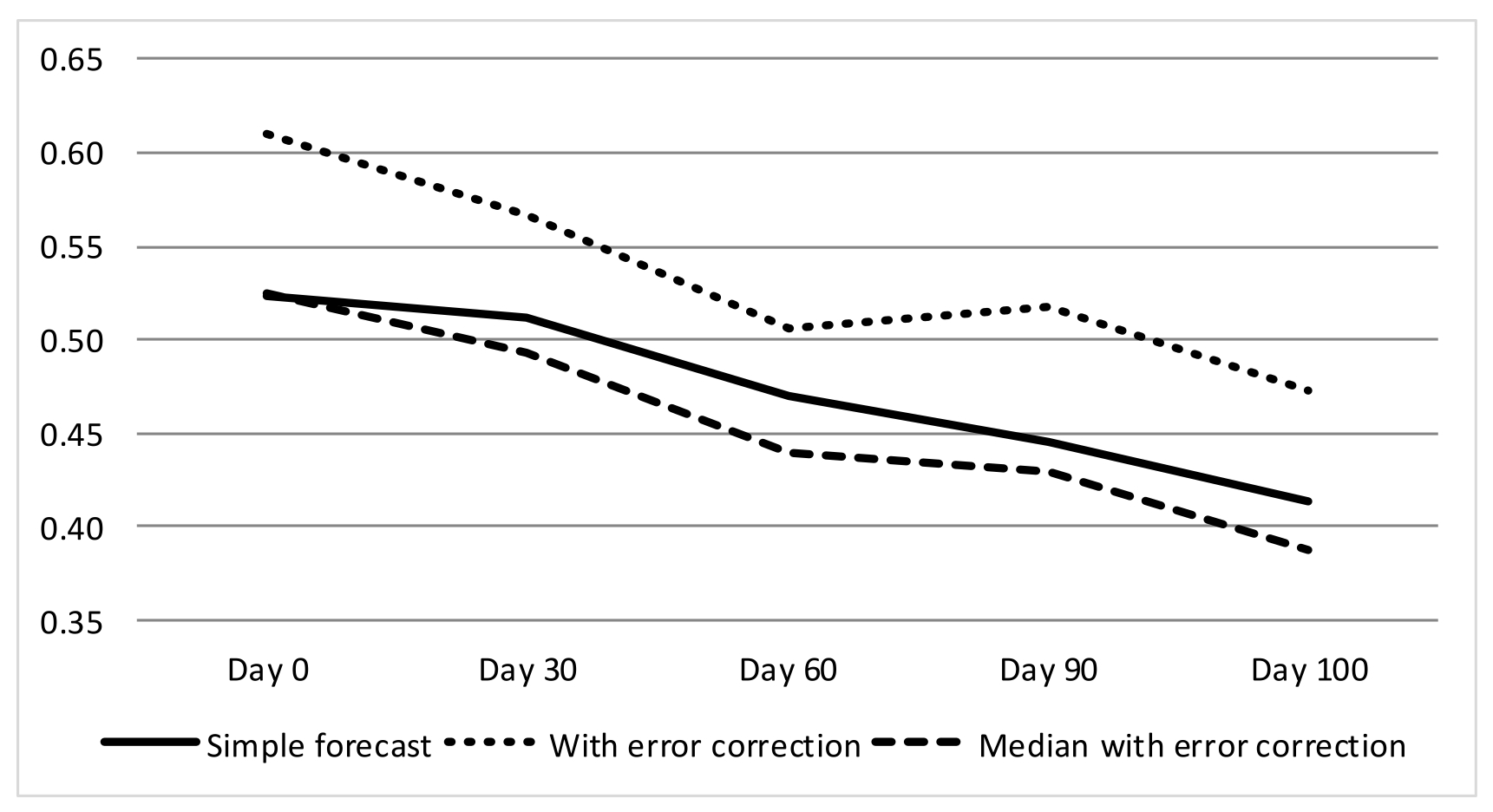}}
	\caption{Mean absolute out-of-sample forecast error of models 3 and 4, with and without last error correction, progressively updated with more data over the current quarter (percentage points, 2002Q1-2019Q4)}
	\label{fig:mod3_4}
\end{figure}

\begin{figure}[!h]
	\centering
	\subfigure[Model 5]{\includegraphics[width=8cm]{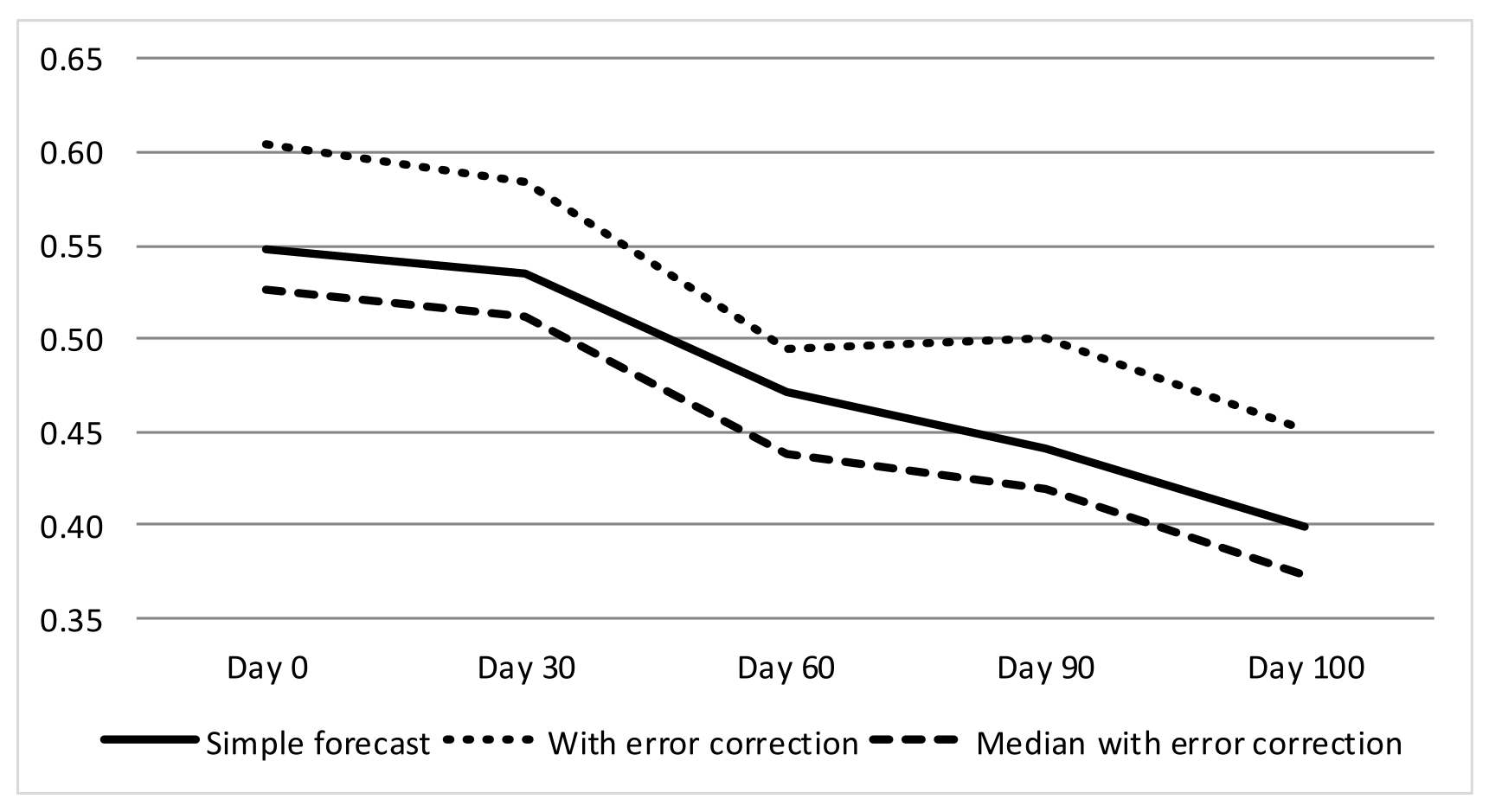}}
	\subfigure[Model 6]{\includegraphics[width=8cm]{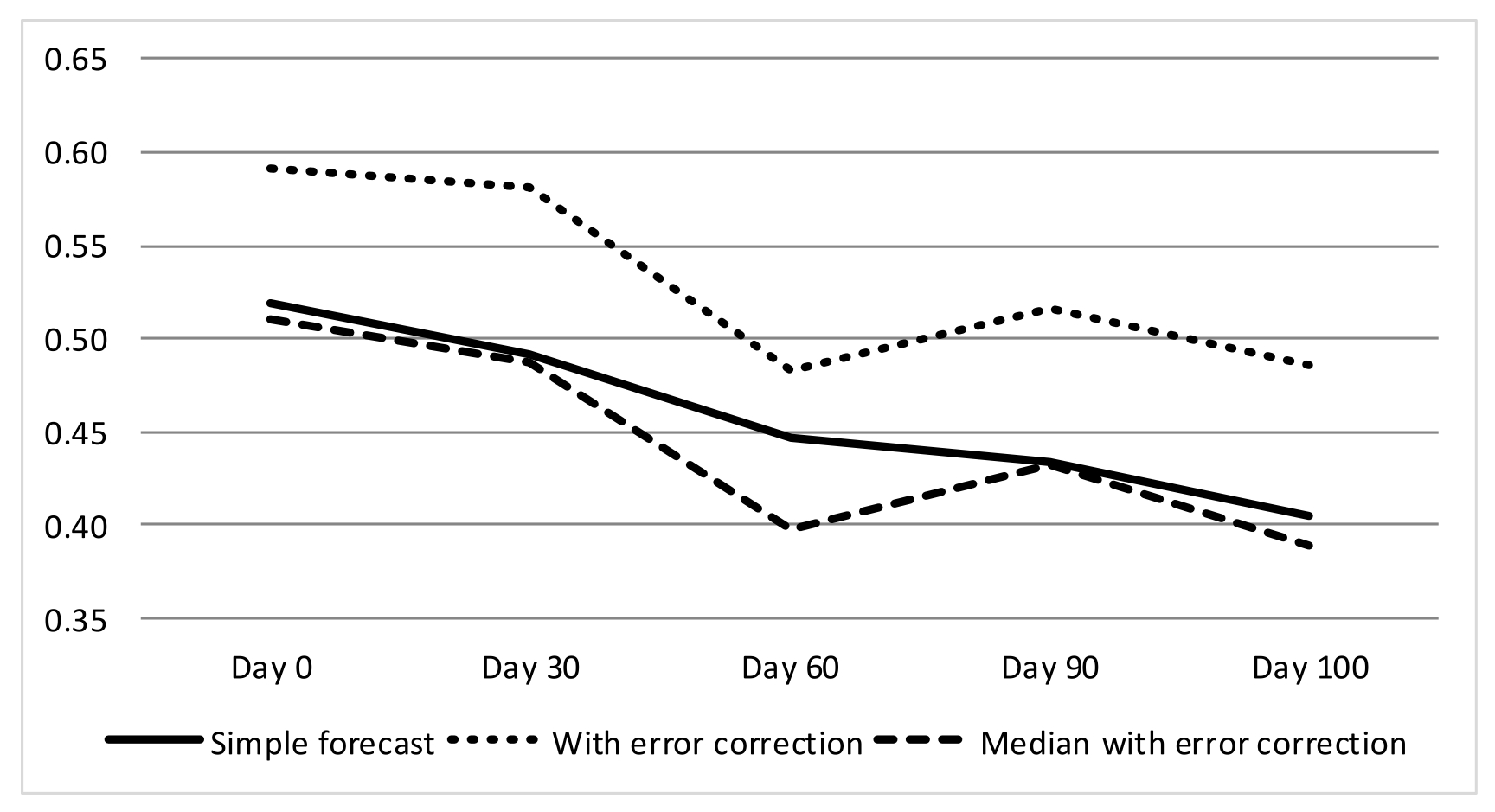}}
	\caption{Mean absolute out-of-sample forecast error of models 5 and 6, with and without last error correction, progressively updated with more data over the current quarter (percentage points, 2002Q1-2019Q4)}
	\label{fig:mod5_6}
\end{figure}

Figures \ref{fig:mod1_2} to \ref{fig:mod5_6} illustrate the out-of-sample mean absolute error (MAE) for each model, progressively updated with more data over the current quarter. For each model and quarter from 2002Q1 to 2019Q4, we computed the direct (simple) forecast without last error correction, the forecast with that kind of correction (as described previously) and the median of these two estimates. A first evidence is that the simple mechanism of picking the last out-of-sample error might not be effective in reducing the MAE, except in some cases for model 3. Nevertheless, the intermediate point between the simple and corrected forecasts always performs better. This is an expected result in the sense that median has been proven very powerful for attenuating or even removing noise in time series \citep{Wen1999}.

A second evidence is that the MAE becomes smaller and smaller from day 0 to day 100, that is, the incorporation of more information concerned with the current quarter improves the quality of the nowcasting exercise. This result was observed in all models and it is particularly evident from day 30 to day 60, recalling that the national accounts (GDP, exports and imports) for the previous quarter become fully available only at day 60 of the current quarter. Thus, that information should be extremely important to improve the accuracy of the GDP growth estimates in addition to the readily available data.

\begin{figure}[!h]
	\centering
	\subfigure[Median of simple forecasts]{\includegraphics[width=8cm]{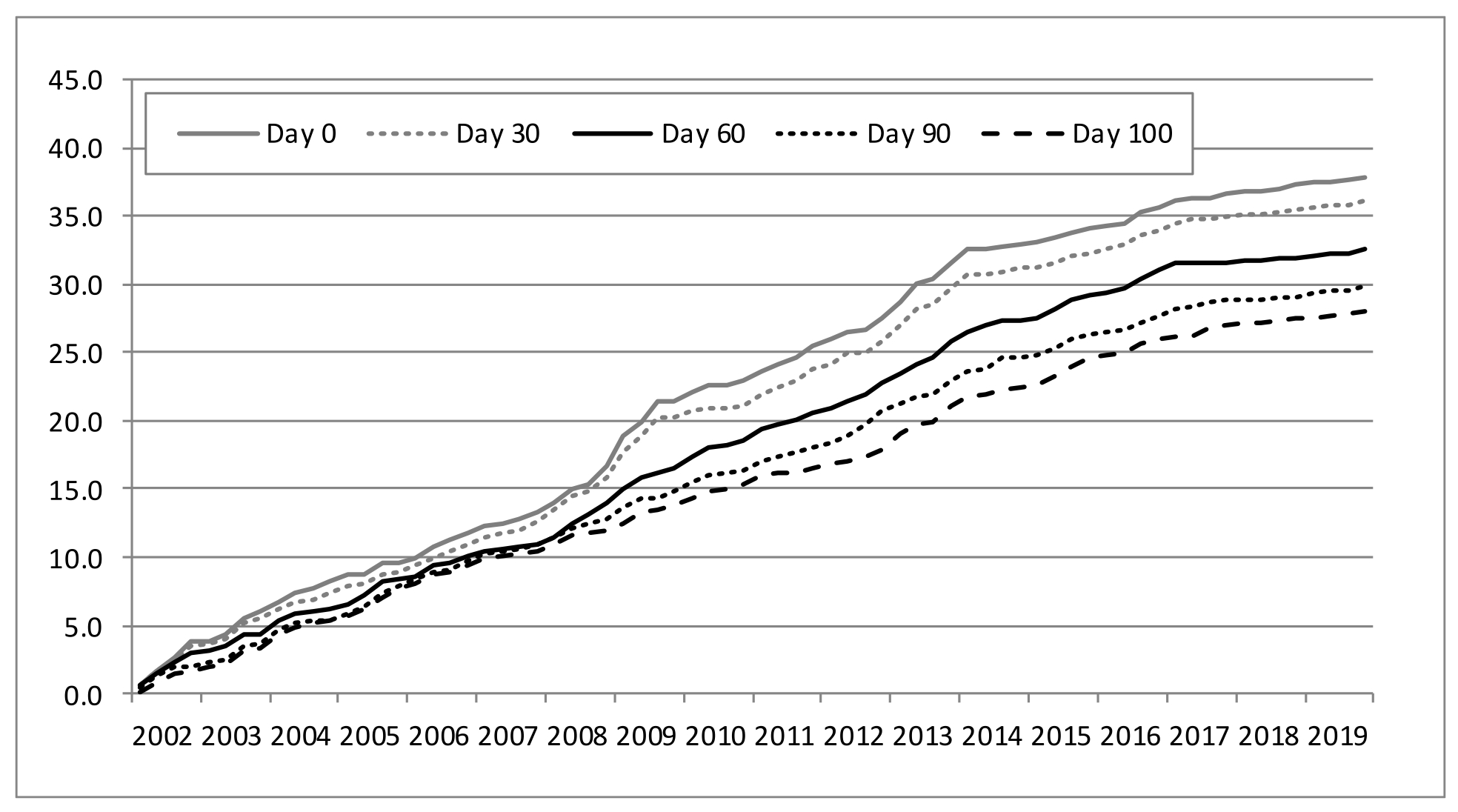}}
	\subfigure[Median with error correction]{\includegraphics[width=8cm]{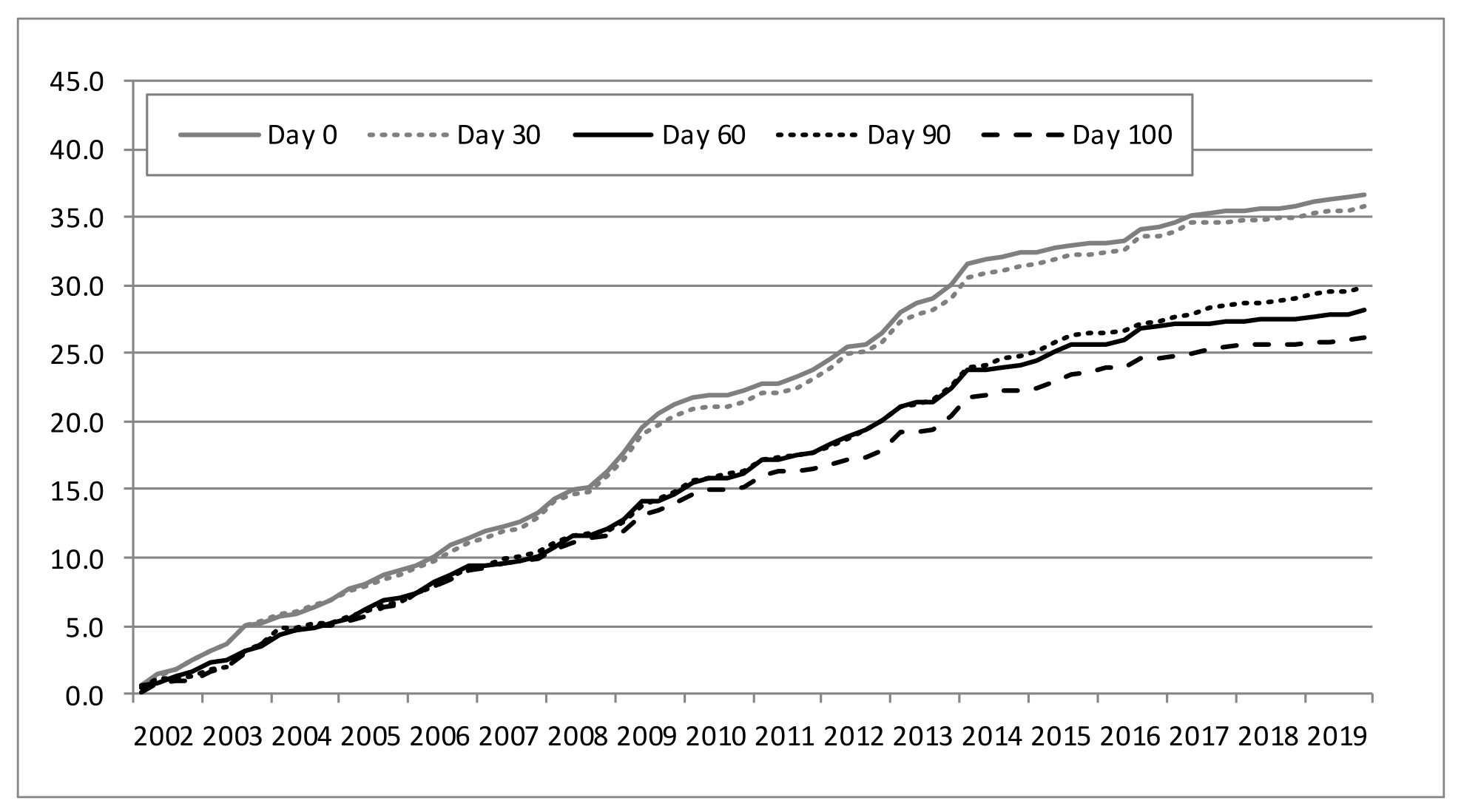}}
	\caption{Cumulative absolute out-of-sample forecast error of the median of models 1 to 6, with and without last error correction, progressively updated with more data over the current quarter (percentage points, 2002Q1-2019Q4)}
	\label{fig:cumerror1}
\end{figure}

Figure \ref{fig:cumerror1} compares the cumulative absolute out-of-sample error of the median of the simple forecasts given by the six models (\ref{eq:consensus}) with the overall median with and without error correction (\ref{eq:consensus_wide}). The reduction of the absolute error by including the last error correction in the median is evident, especially for day 60 and beyond. Additionally, this figure confirms the relevance of using, progressively, more and more data over the current quarter. In fact, monthly data are still important in the sense that the cumulative absolute error reduces from day zero to 30 and even more from day 60 to days 90 and 100. Our forecasts are being published typically at day 100 with a sightly error reduction from day 90.

The gain in terms of out-of-sample MAE between day zero and day 60 is 0.08 percentage points (pp) from 0.53 to 0.45, as suggested by the last column of table \ref{tab:results}. With last error correction, the gain is slightly better, 0.09 pp, from 0.51 to 0.42, see table \ref{tab:results_corr}. The inclusion of more high-frequency data concerned with the current quarter can improve the MAE additionally in 0.06 pp towards a final mark of 0.36 for day 100 with last error correction. This kind of improvement is also visible in mean squared error (MSE) and its root, which is directly comparable with MAE.

\begin{figure}[!h]
	\centering
	\subfigure[Median of simple forecasts]{\includegraphics[width=8cm]{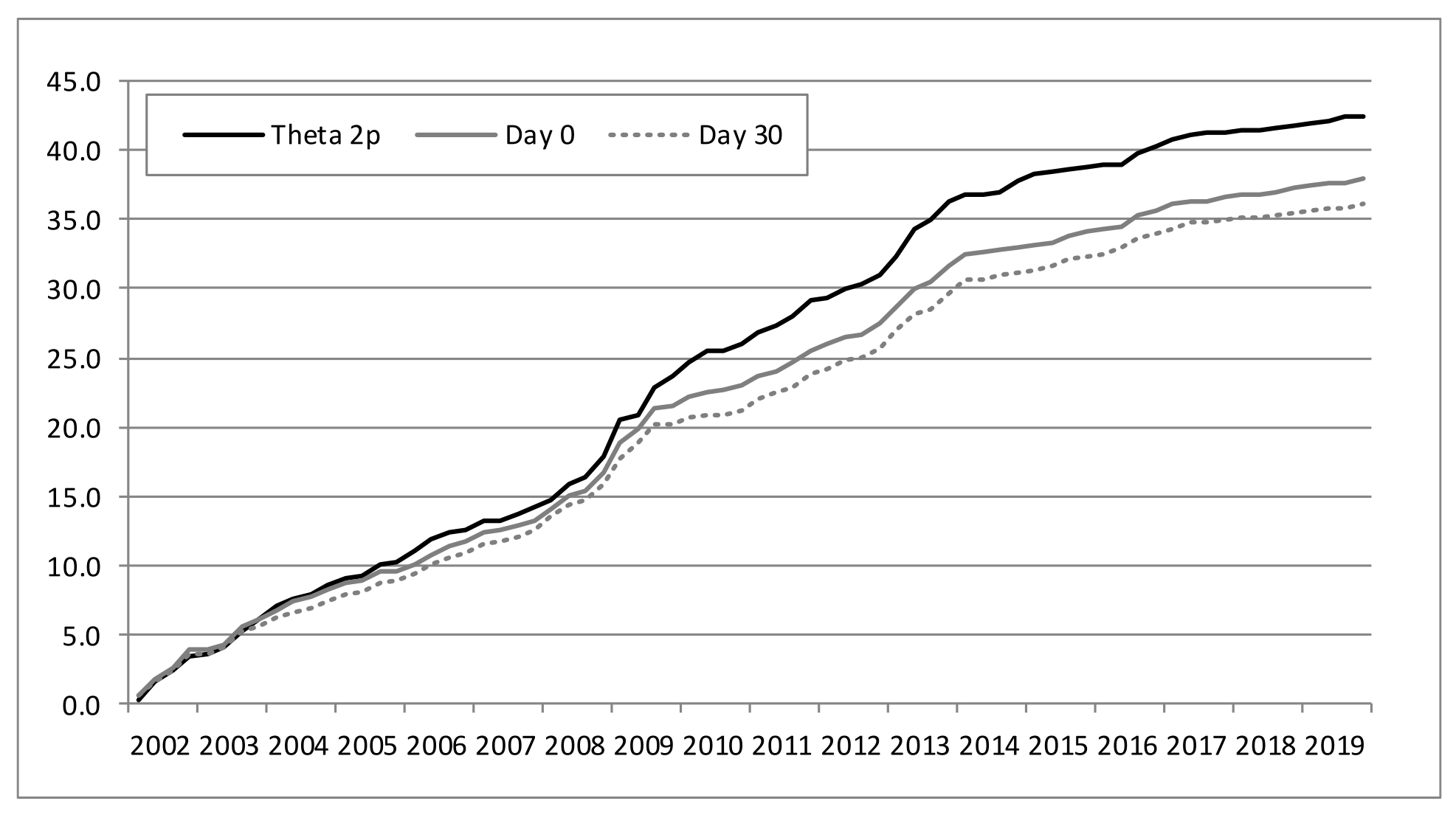}}
	\subfigure[Median with error correction]{\includegraphics[width=8cm]{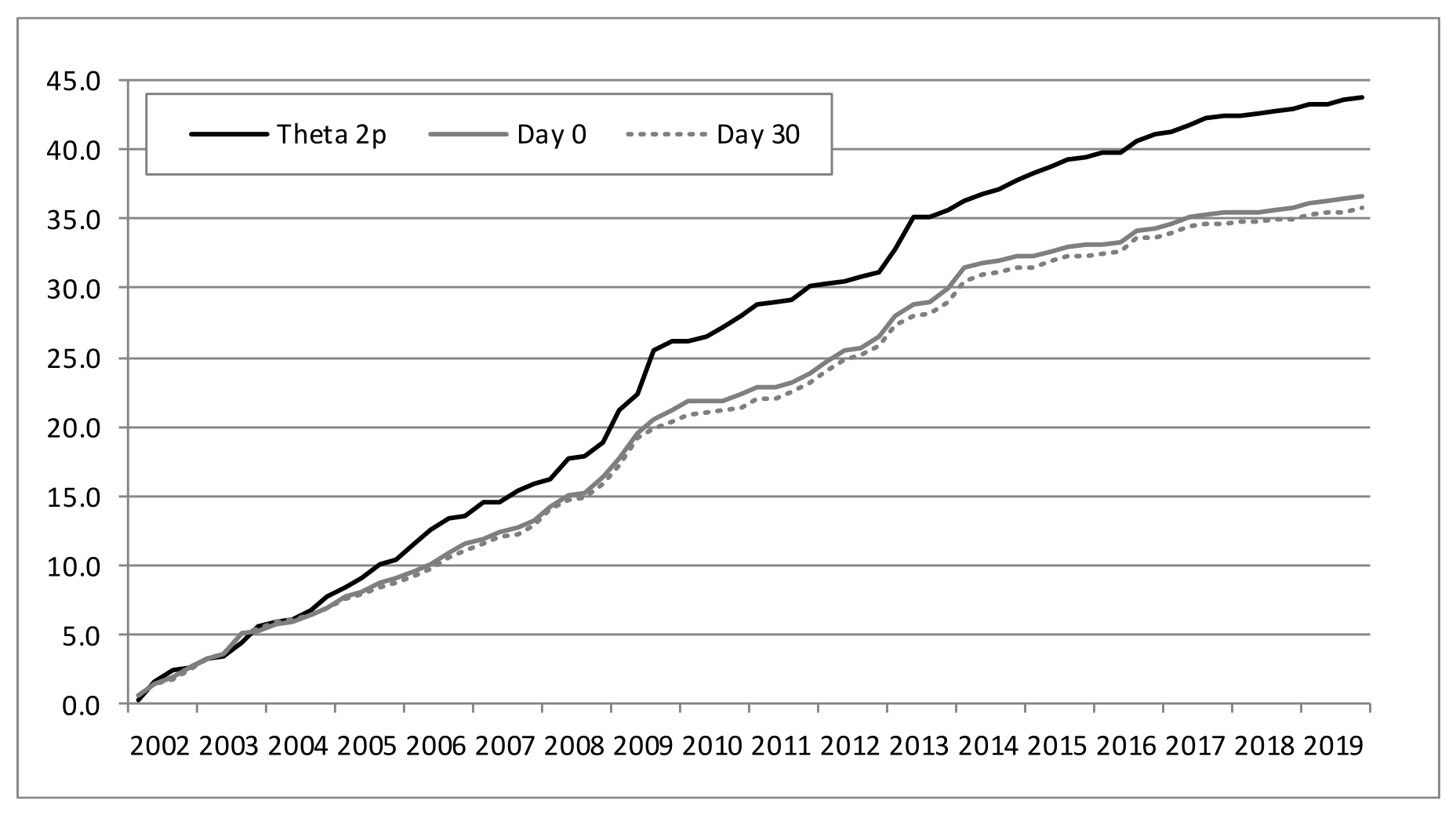}}
	\caption{Cumulative absolute out-of-sample forecast error of the median of models 1 to 6, with and without last error correction, given the data available at days 0 and 30 of the current quarter, compared with the 2 step-ahead Theta forecast  (percentage points, 2002Q1-2019Q4)}
	\label{fig:cumerror2}
\end{figure}

\begin{figure}[!h]
	\centering
	\subfigure[Median of simple forecasts]{\includegraphics[width=8cm]{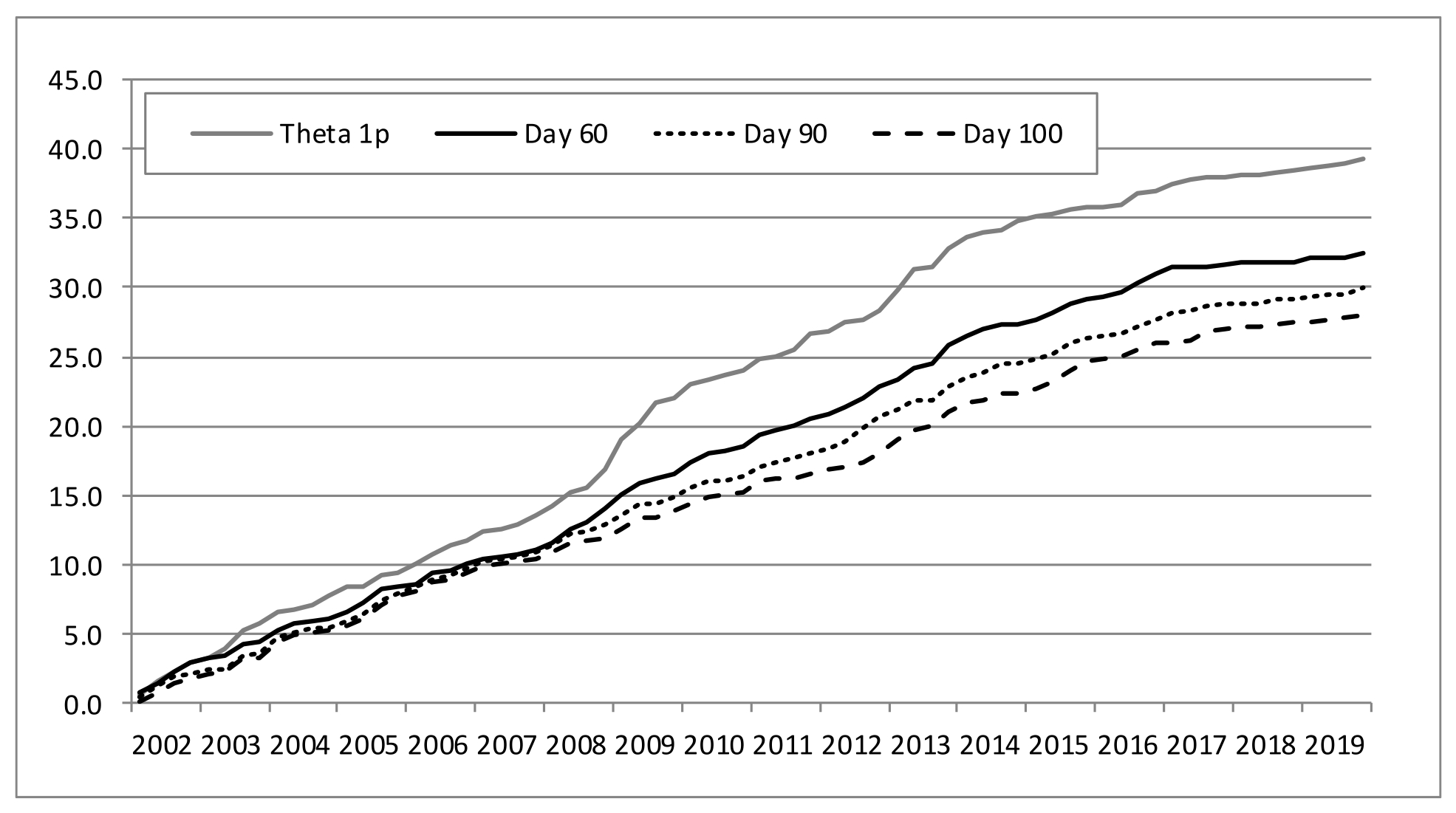}}
	\subfigure[Median with error correction]{\includegraphics[width=8cm]{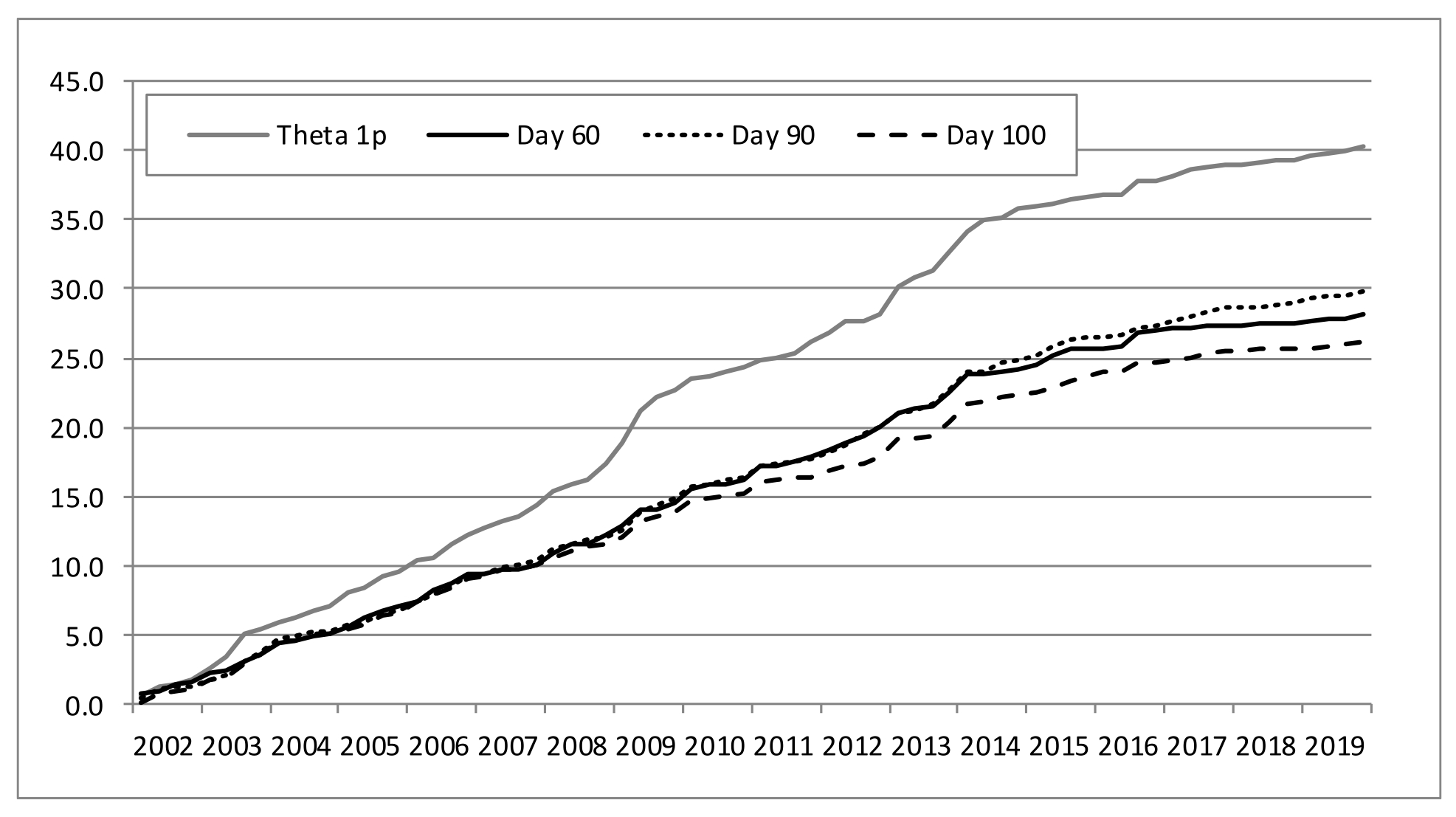}}
	\caption{Cumulative absolute out-of-sample forecast error of the median of the models 1 to 6, with and without last error correction, given the data available at days 60, 90 and 100 of the current quarter, compared with the 1 step-ahead Theta forecast  (percentage points, 2002Q1-2019Q4)}
	\label{fig:cumerror3}
\end{figure}

As suggested by figures \ref{fig:cumerror2} and \ref{fig:cumerror3} (and the same tables), our approach performs quite better than the Theta method. We found also that the last correction error may increase the MAE of the estimates computed with that method. Thus, error correction mechanisms such the one employed here should be avoided, or used with care, as far as the Theta method is concerned.

The Targeted Diffusion Index (TDI) model estimated by \citet{Dias2016} used a comparable dataset but for a shorter period (2002Q1-2015Q4). For this sub-sample, our approach gave a MAE of 0.42 pp which is close to 0.41 from TDI model. In root MSE, the difference between the two methods is meaningless at days 60 and 90 and our approach may perform better at day 100 (see table \ref{tab:results_sub}).

\section{Conclusion}
\label{sec:conclusion}

This article shows how relatively simple it is to design robust statistical models to nowcast macroeconomic variables of interest, like GDP, based on readily available indicators. To this end, it is important to mobilize either qualitative or quantitative high-frequency indicators correlated with GDP. In the former case, we have the coincident indicators of business sentiment or economic climate, and in the second case, quantitative indicators like industrial production, cement and car sales, or the value of exports of goods and services.

Both types of indicators (qualitative and quantitative) can be aggregated quarterly and used in simple multivariate linear regressions in the year-on-year (or chain) variation of GDP. In this quarterly aggregation, the monthly jagged values not yet available for the current quarter can be estimated in a very simple way, using a tendency-cycle decomposition such as the proposed by \citet{Hodrick1997}, correcting any excessive noise with moving averages and using the median to get consensus among different estimates, eventually corrected from the last observed forecast error.

In fact, these rather practical procedures, although laborious and demanding, can lead to predictions with an average error similar to the one associated with more sophisticated methods like the Targeted Diffusion Index (TDI) which involves the treatment of tens or even hundreds of time series. The method described throughout the article has proved to be  effective against the recognized Theta method, which is particularly good in its simplicity, and it can be replicated to a country rather than Portugal straightforwardly.

\section*{Acknowledgments}
\label{sec:acknowledgments}

This work was supported by Fundação para a Ciência e Tecnologia (FCT), Lisbon, Portugal under a post-doctoral grant with the reference CUBE-MACROECO-BPD4 from Católica Lisbon Research Unit in Business \& Economics (UID/GES/00407/2020).

\bibliography{library}

\begin{thebibliography}{15}
\providecommand{\natexlab}[1]{#1}
\providecommand{\url}[1]{\texttt{#1}}
\expandafter\ifx\csname urlstyle\endcsname\relax
  \providecommand{\doi}[1]{doi: #1}\else
  \providecommand{\doi}{doi: \begingroup \urlstyle{rm}\Url}\fi

\bibitem[Assimakopoulos and Nikolopoulos(2000)]{Assimakopoulos2000}
V.~Assimakopoulos and K.~Nikolopoulos.
\newblock The theta model: a decomposition approach to forecasting.
\newblock \emph{International Journal of Forecasting}, 16:\penalty0 521--530,
  2000.
\newblock URL
  \url{https://www.researchgate.net/publication/223049702_The_theta_model_A_decomposition_approach_to_forecasting}.

\bibitem[Bai and Ng(2008)]{Bai2008}
J.~Bai and S.~Ng.
\newblock Forecasting economic time series using targeted predictors.
\newblock \emph{Journal of Econometrics}, 2:\penalty0 304--317, 2008.

\bibitem[Dias et~al.(2010)Dias, Pinheiro, and Rua]{Dias2010}
F.~Dias, M.~Pinheiro, and A.~Rua.
\newblock Forecasting using targeted diffusion indexes.
\newblock \emph{Journal of Forecasting}, 29:\penalty0 341--352, 2010.
\newblock \doi{10.1002/for.1132}.

\bibitem[Dias et~al.(2015)Dias, Pinheiro, and Rua]{Dias2015}
F.~Dias, M.~Pinheiro, and A.~Rua.
\newblock Forecasting portuguese gdp with factor models: Pre- and post-crisis
  evidence.
\newblock \emph{Economic Modelling}, 44:\penalty0 266--272, 2015.

\bibitem[Dias et~al.(2016)Dias, Pinheiro, and Rua]{Dias2016}
F.~Dias, M.~Pinheiro, and A.~Rua.
\newblock Previsão do pib através de uma abordagem bottom-up num contexto
  rico em informação.
\newblock \emph{Revista de Estudos Económicos}, II:\penalty0 1--20, 2016.

\bibitem[Giannone et~al.(2008)Giannone, Reichlin, and Small]{Giannone2008}
D.~Giannone, L.~Reichlin, and D.~Small.
\newblock Nowcasting: The real-time informational content of macroeconomic
  data.
\newblock \emph{Journal of Monetary Economics}, 55:\penalty0 665--676, 2008.

\bibitem[Higgins(2014)]{Higgins2014}
P.~Higgins.
\newblock Gdpnow: A model for gdp “nowcasting”.
\newblock \emph{Federal Reserve Bank of Atlanta Working Paper}, 7, 2014.
\newblock URL
  \url{https://www.frbatlanta.org/research/publications/wp/2014/07.aspx}.

\bibitem[Hodrick and Prescott(1997)]{Hodrick1997}
R.~J. Hodrick and E.~C. Prescott.
\newblock Postwar u.s. business cycles: An empirical investigation.
\newblock \emph{Journal of Money, Credit and Banking}, 29:\penalty0 1--16,
  1997.
\newblock ISSN 00222879.
\newblock \doi{10.2307/2953682}.
\newblock URL
  \url{https://www0.gsb.columbia.edu/faculty/rhodrick/prescott-hodrick1997.pdf}.

\bibitem[Hyndman and Billah(2001)]{Hyndman2001}
R.~J. Hyndman and B.~Billah.
\newblock Unmasking the theta method.
\newblock \emph{Department of Econometrics and Business Statistics, Monash
  University}, 27.09:\penalty0 1--7, 2001.
\newblock URL \url{https://robjhyndman.com/papers/Theta.pdf}.

\bibitem[Klein and Sojo(1989)]{Klein1989}
L.~R. Klein and E.~Sojo.
\newblock Combinations of high and low frequency data in macroeconometric
  models.
\newblock \emph{Economics in Theory and Practice: An Eclectic Approach}, pages
  3--16, 1989.

\bibitem[Miller and Chin(1996)]{Miller1996}
P.~J. Miller and D.~M. Chin.
\newblock Using monthly data to improve quarterly model forecasts.
\newblock \emph{Federal Reserve Bank of Minneapolis Quarterly Review},
  20:\penalty0 16--33, 1996.

\bibitem[Stock and Watson(1989)]{Stock1989}
J.~H. Stock and M.~W. Watson.
\newblock New indexes of coincident and leading economic indicators, 1989.
\newblock URL \url{http://www.nber.org/chapters/c10968}.

\bibitem[Stock and Watson(2002)]{Stock2002}
J.~H. Stock and M.~W. Watson.
\newblock Macroeconomic forecasting using diffusion indexes.
\newblock \emph{Journal of Business and Economic Statistics}, 20:\penalty0
  147--162, 2002.

\bibitem[Stock and Watson(2011)]{Stock2011}
J.~H. Stock and M.~W. Watson.
\newblock Dynamic factor models, 2011.
\newblock URL
  \url{http://www.oxfordhandbooks.com/view/10.1093/oxfordhb/9780195398649.001.0001/oxfordhb-9780195398649}.

\bibitem[Wen and Zeng(1999)]{Wen1999}
Y.~Wen and B.~Zeng.
\newblock A simple nonlinear filter for economic time series analysis.
\newblock \emph{Economic Letters}, 64:\penalty0 151--160, 1999.
\newblock URL
  \url{https://www.sciencedirect.com/science/article/abs/pii/S0165176599000890}.

\end{thebibliography}

\pagebreak

\section*{Annex}
\label{sec:annex}

\begin{table}[!h]
	\centering
	\caption{Estimated coefficients (full sample 1996Q1 - 2019Q4)}
	\label{tab:coef_2019}
	\begin{tabular}{lcccccc}

		\\[-2ex]\hline

		\multicolumn{1}{}{} & \multicolumn{6}{c}{Model} \\
		\cline{2-7}

		Coefficients & 1		& 2				& 3				& 4				& 5				& 6 \\

		\hline

		sum		& 0.497$^{***}$	& 0.477$^{***}$	& 0.545$^{***}$	& 0.458$^{***}$	& 0.448$^{***}$	& 0.525$^{***}$	\\ 
				& (0.080)		& (0.064)		& (0.087)		& (0.080)		& (0.069)		& (0.080)		\\ 
		
		ESI-100	& 0.055$^{***}$	& 0.020			& 0.110$^{***}$	& 0.108$^{***}$	& 0.088$^{***}$	& 0.039$^{**}$	\\ 
				& (0.016)		& (0.014)		& (0.013)		& (0.011)		& (0.011)		& (0.018)		\\

		ICE		& 0.310$^{***}$	& 0.366$^{***}$	&				&				&				& 0.312$^{***}$	\\ 
				& (0.060)		& (0.050)		&				&				&				& (0.059)		\\

		ipi		& 0.068$^{***}$	& 0.015			& 0.069$^{***}$	& 0.076$^{***}$	& 0.034$^{**}$	& 0.055$^{***}$	\\ 
				& (0.016)		& (0.015)		& (0.018)		& (0.016)		& (0.015)		& (0.017)		\\

		cem		& 0.018$^{***}$	& 0.015$^{***}$	& 0.025$^{***}$	& 0.019$^{***}$	& 0.017$^{***}$	& 0.018$^{***}$	\\ 
				& (0.005)		& (0.004)		& (0.005)		& (0.005)		& (0.004)		& (0.005)		\\

		car		&        		&          		& 0.010$^{**}$ 	&      			& 				&				\\
        		&              	&           	& (0.004)     	&      			& 				&				\\

		atm		&              	&	            &				& 0.071$^{***}$	& 0.071$^{***}$	&			 	\\
        		&				&               &               & (0.013) 		& (0.012)		&	 			\\

		exp     &              	& 0.042$^{***}$	&               &      			& 0.037$^{***}$	&				\\
	        	&              	& (0.013)       &               &      			& (0.014) 		&				\\

		imp		&              	& 0.063$^{***}$	&               &      			& 0.051$^{***}$	&				\\
				&              	& (0.014)		&				&      			& (0.015)		&	 			\\

		CEPR	&        		&              	&				&      			& 				& 0.662$^{***}$	\\ 
        		&              	&               &				&      			& 				& (0.107)		\\

		Constant& 0.736$^{***}$	& 0.269			& 0.937$^{***}$	& 0.457$^{***}$	& 0.095			& 0.381$^{***}$	\\ 
				& (0.092)  		& (0.098)		& (0.090)		& (0.122)		& (0.124)		& (0.132)		\\		
		
		\hline
		
		Observations		& 96		& 96			& 96			& 96			& 96			& 96	\\ 

		R$^{2}$\			& 0.948		& 0.968			& 0.938			& 0.950			& 0.964			& 0.951	\\ 

		Adjusted R$^{2}$	& 0.946		& 0.965			& 0.934			& 0.947			& 0.961			& 0.947	\\ 

		Residual Std. Error			& 0.556		& 0.443			& 0.611			& 0.549			& 0.472			& 0.547	\\ 

		F statistic	& 331.1$^{***}$ & 379.8$^{***}$ & 271.7$^{***}$ & 339.7$^{***}$ & 333.9$^{***}$ & 285.9$^{***}$	\\ 
		
		\hline

		\textit{p-values:}	&	&	&	& $^{*}p<$0.1;	& $^{**}p<$0.05;	& $^{***}p<$0.01	\\
		
	\end{tabular}
\end{table}

\begin{table}[p]
	\centering
	\caption{Estimated coefficients (sub-sample 1996Q1 - 2005Q4)}
	\label{tab:coef_2005}
	\begin{tabular}{lcccccc}

		\\[-2ex]\hline

		\multicolumn{1}{}{} & \multicolumn{6}{c}{Model} \\
		\cline{2-7}

		Coefficients & 1		& 2				& 3				& 4				& 5				& 6 \\

		\hline

  		sum		& 0.512$^{***}$	& 0.470$^{***}$	& 0.503$^{***}$	& 0.561$^{***}$	& 0.550$^{***}$	& 0.522$^{***}$	\\ 
				& (0.167)		& (0.140)		& (0.157)		& (0.169)		& (0.156)		& (0.170)		\\ 

		ESI-100	& 0.029			& -0.007			& 0.090$^{***}$	& 0.091$^{***}$	& 0.058$^{**}$	& 0.020		\\ 
				& (0.030)		& (0.027)		& (0.020)		& (0.022)		& (0.024)		& (0.036)		\\

		ICE		& 0.388$^{***}$	& 0.375$^{***}$	&				&				&				& 0.398$^{***}$	\\ 
				& (0.123)		& (0.108)		&				&				&				& (0.125)		\\

		ipi		& 0.023			& 0.029			& 0.032			& 0.040			& 0.049			& 0.027			\\ 
				& (0.032)		& (0.027)		& (0.029)		& (0.032)		& (0.030)		& (0.034)		\\

		cem		& 0.018$^{***}$	& 0.015$^{**}$	& 0.023$^{***}$	& 0.019$^{***}$	& 0.020$^{***}$	& 0.018$^{***}$	\\ 
				& (0.006)		& (0.006)		& (0.005)		& (0.006)		& (0.006)		& (0.006)		\\

		car		&        		&              	& 0.031$^{***}$ &      			& 				&				\\
        		&              	&               & (0.008)      	&      			& 				&				\\

		atm 	&	            &		        &				& 0.066$^{**}$	& 0.038			&			 	\\
        		&				&               &               & (0.024) 		& (0.025)		&	 			\\

		exp     &              	& 0.020			&               &      			& 0.035			&				\\
	        	&              	& (0.031)       &               &      			& (0.036) 		&				\\

		imp		&             	& 0.092$^{***}$	&               &      			& 0.073$^{**}$	&				\\
				&              	& (0.025)		&				&      			& (0.029)		&	 			\\

		CEPR	&        		&              	&				&      			& 				& 0.200			\\ 
        		&              	&               &				&      			& 				& (0.463)		\\

		Constant& 0.6632$^{***}$& 0.302			& 1.348$^{***}$	& 0.465			& 0.393			& 0.569			\\ 
				& (0.233)  		& (0.216)		& (0.179)		& (0.307)		& (0.284)		& (0.321)		\\

		\hline

		Observations	& 40		& 40			& 40			& 40			& 40			& 40		\\ 

		R$^{2}$			& 0.918		& 0.946			& 0.927			& 0.913			& 0.931			& 0.919		\\ 

		Adjusted R$^{2}$& 0.906		& 0.934			& 0.916			& 0.900			& 0.915			& 0.904		\\ 

		Residual Std. Error	& 0.594	& 0.499			& 0.562			& 0.613			& 0.565			& 0.601		\\ 

		F statistic		& 76.5$^{***}$ & 79.9$^{***}$ & 82.2$^{***}$ & 71.6$^{***}$ & 62.2$^{***}$ & 62.3$^{***}$	\\ 

		\hline

		\textit{p-values:}	&	&	&	& $^{*}p<$0.1;	& $^{**}p<$0.05;	& $^{***}p<$0.01	\\
		
	\end{tabular}
\end{table}

\begin{table}[p]
	\centering
	\caption{Mean squared and absolute out-of-sample forecast errors of the median of the models (1) to (6) conditioned to available data over the current quarter (full sample 2002Q1-2019Q4)}
	\label{tab:results}

	\begin{tabular}{lccc}
		\\[-2ex]\hline
		Data available at: & Mean Squared Error & Root MSE & Mean Absolute Error \\
		\hline
		Day 0								& 0.45	& 0.67	& 0.53 \\
		Day 30								& 0.40	& 0.63	& 0.50 \\
		\emph{Benchmark:} Theta model 2p         & 0.60	& 0.77	& 0.59 \\
		\hline
		Day 60								& 0.30	& 0.54	& 0.45 \\
		Day 90								& 0.25	& 0.50	& 0.42 \\
		Day 100								& 0.23	& 0.48	& 0.39 \\
		\emph{Benchmark:} Theta model 1p	      & 0.49	& 0.70	& 0.54 \\
		\hline
	\end{tabular}

\end{table}

\pagebreak

\begin{table}[h]
	\centering
	\caption{Mean squared and absolute out-of-sample forecast errors of the median of the models (1) to (6) conditioned to available data over the current quarter, after last error correction (full sample 2002Q1-2019Q4)}
	\label{tab:results_corr}

	\begin{tabular}{lccc}
		\\[-2ex]\hline
		Data available at: & Mean Squared Error & Root MSE & Mean Absolute Error \\
		\hline
		Day 0								& 0.42	& 0.65	& 0.51 \\
		Day 30								& 0.41	& 0.64	& 0.50 \\
		\emph{Benchmark:} Theta model 2p	      & 0.69	& 0.83	& 0.61 \\
		\hline
		Day 60								& 0.26	& 0.51	& 0.42 \\
		Day 90								& 0.26	& 0.51	& 0.42 \\
		Day 100								& 0.23	& 0.48	& 0.36 \\
		\emph{Benchmark:} Theta model 1p	      & 0.52	& 0.72	& 0.56 \\
		\hline
	\end{tabular}

\end{table}

\begin{table}[h]
	\centering
	\caption{Mean squared and absolute out-of-sample forecast errors of the median of the models (1) to (6) conditioned to available data over the current quarter, after last error correction (sub-sample 2002Q1-2015Q4)}
	\label{tab:results_sub}

	\begin{tabular}{lccc}
		\\[-2ex]\hline
		Data available at: & Mean Squared Error & Root MSE & Mean Absolute Error \\
		\hline
		Day 0								& 0.52	& 0.72	& 0.59 \\
		Day 30								& 0.49	& 0.70	& 0.58 \\
		\emph{Benchmark:} Theta model 2p	      & 0.85	& 0.92	& 0.70 \\
		\hline
		Day 60								& 0.32	& 0.56	& 0.46 \\
		Day 90								& 0.32	& 0.56	& 0.47 \\
		Day 100								& 0.28	& 0.53	& 0.42 \\
		\emph{Benchmark:} Theta model 1p	      & 0.64	& 0.80	& 0.66 \\
		\emph{Benchmark:} TDI model			& 0.30	& 0.55	& 0.41 \\
		\hline
	\end{tabular}

\end{table}

\pagebreak

\end{document}